\documentclass[a4paper,fleqn,usenatbib]{mn2e}
\usepackage{color}
\usepackage{epsfig}
\usepackage{epstopdf}
\usepackage{amsmath, amssymb}
\usepackage{hyperref}
\usepackage{wasysym} 
\usepackage{multirow}
\usepackage{lipsum}
\usepackage{ulem}
\usepackage{setspace}
\usepackage{float}
\usepackage[T1]{fontenc}
\usepackage{aecompl}
\usepackage{graphicx,color,bm}
\usepackage{latexsym}
\usepackage{epsfig}
\usepackage{amssymb}
\usepackage{amsmath}
\usepackage{wasysym}
\usepackage{pstricks}
\usepackage{enumitem}

\def\etal{{\frenchspacing\it et al.}}

\def\be{\begin{equation}}
\def\ee{\end{equation}}
\def\ba{\begin{eqnarray}}
\def\ea{\end{eqnarray}}

\newcommand{\llan}{\left\langle}
\newcommand{\rran}{\right\rangle}

\def\LaTeX{L\kern-.36em\raise.3ex\hbox{a}\kern-.15em
    T\kern-.1667em\lower.7ex\hbox{E}\kern-.125emX}

\def\vc#1{{\bf#1}}

\begin{document}

\voffset-1.25cm
\title[Testing a new approach to measure the evolution of the structure growth]{The extended Baryon Oscillation Spectroscopic Survey (eBOSS):
testing a new approach to measure the evolution of the structure growth} 
\author[Ruggeri \etal]{
\parbox{\textwidth}{
Rossana Ruggeri$^{1}$\thanks{Email: rossana.ruggeri@port.ac.uk},  Will J. Percival $^{1}$,   Eva-Maria Mueller $^{1}$,  H\'ector Gil-Mar\'in $^{2,3}$,   Fangzhou Zhu $^{4}$, Nikhil Padmanabhan$^{4}$, Gong-Bo Zhao $^{5,1}$ 
}
\vspace*{15pt} \\
$^1$ Institute of Cosmology \& Gravitation, University of Portsmouth, Dennis Sciama Building, Portsmouth, PO1 3FX, UK\\
$^2$ Sorbonne Universit\'e, Institut Lagrange de Paris (ILP), 98 bis Boulevard Arago, 75014 Paris, France \\
$^3$ Laboratoire de Physique Nucl\'eaire et de Hautes Energies, Universit\'e Pierre et Marie Curie, Paris, France \\
$^4$ Dept. of Physics, Yale University, New Haven, CT 06511\\
$^5$ National Astronomy Observatories, Chinese Academy of Science, Beijing, 100012, P.R.China
}
\date{\today} 
\pagerange{\pageref{firstpage}--\pageref{lastpage}}

\label{firstpage}

\maketitle


\begin{abstract}
  The extended Baryon Oscillation Spectroscopic Survey (eBOSS) is one
  of the first of a new generation of galaxy redshift surveys that
  will cover a large range in redshift with sufficient resolution to
  measure the baryon acoustic oscillations (BAO) signal. For surveys covering a large redshift range
  we can no longer ignore cosmological evolution, meaning that either
  the redshift shells analysed have to be significantly narrower than the survey, or we have to allow
  for the averaging over evolving quantities. Both of these have the
  potential to remove signal: analysing small volumes increases the
  size of the Fourier window function, reducing the large-scale
  information, while averaging over evolving quantities can, if not performed carefully, remove
  differential information.
   It will be important to  measure cosmological evolution from these surveys to explore and
  discriminate between models.  We apply a method to optimally extract
  this differential information to mock catalogues designed to mimic
  the eBOSS quasar sample. By applying a set of weights to extract
  redshift space distortion measurements as a function of redshift, we
  demonstrate an analysis that does not invoke the problems discussed
  above. We show that our estimator gives unbiased constraints.
\end{abstract}


\section{INTRODUCTION} 
\label{intro}

The eBOSS survey \citep{2016dawson, 2016Gongbo, 2017blanton}, which commenced in
July 2014, will cover the largest volume to date of any cosmological
redshift survey with a density sufficient to extract useful
cosmological information. eBOSS observations will target multiple
density-field tracers, including more than $250,000$ luminous red
galaxies (LRGs), $195,000$ emission line galaxies (ELGs) at effective
redshifts $z = 0.72, 0.87$ and over $500,000$ quasars between
$0.8< z< 2.2$.  The survey's goals include the distance measurement at
$1$$-2\%$ accuracy with the BAO peak on
the LRG sample and the first BAO measurements using quasars as density
tracers over the redshift range
$1<z<2$ (the first clustering measurements were recently presented in
\citet{2017ata}). The wide redshift range covered, compared with
that in previous redshift surveys represents an unique opportunity to
test and discriminate between different cosmological scenarios on the
basis of their evolution in redshift. Full survey details can be found
in \citet{2016dawson}.

The clustering analysis strategy adopted for most recent galaxy survey
analyses was based on computing the correlation function or the power
spectrum for individual samples or subsamples, overwhich the
parameters being measured were assumed to be unvarying with redshift. The
measurements were then considered to have been made at an effective
redshift: see e.g. \cite{Alam2016}, \cite{anderson2011}. In
particular \cite{Alam2016}, divided the full The Baryon Oscillation Spectroscopic Survey (BOSS) survey volume in
three overlapping redshift bins and repeated the measurement in each
sub-volume. This technique has many disadvantages: the choice of bins
is a balance between having enough data for a significant detection in each bin
leading to Gaussian errors and having bins small enough that there is
no cosmological evolution across them, leading to a degrading
compromise. The technique also ignores the correlation between
galaxies in different redshift bins leading to a lower signal to noise
ratio, which in Fourier space is equivalent to having a wider window
function.

To complicate analyses further, many mock catalogues currently used to
compare to the data intrinsically lack evolution, or ``lightcone''
effects, being drawn from simulation snapshots. Although this is a
separate problem, these differences limit the tests of the effects of
evolution that can be performed, and have the potential to hide biases
caused by evolution.

Recent work by \cite{zhu2014}, \cite{2016Mio}, and \cite{2017eva}
introduced an alternative approach to the redshift binning. The idea
is to consider the whole volume of the survey and optimally compress
the information in the redshift direction by applying a set of
redshift weights to all galaxies, and only then computing the weighted
correlation function. Comparing measurements made using different sets
of redshift weights maintains the sensitivity to the underlying
evolving theory. The sets of weights are derived in order to minimize
the error on the parameters of interest. In addition, by applying the
redshift weighting technique instead of splitting the survey, is it
possible to compute the correlation function to larger scales whilst
accounting for the evolution in redshift; this was particularly clear
in \cite{2017eva}, which considered  this method to optimize the
measurement of local primordial non gaussianity, which relies on large
scales.  Further, \cite{2016Zhu} showed that the
application of a weighting scheme rather than splitting into bins
also improves BAO measurements.

The need to correctly deal with evolution will increase for the DESI
and Euclid experiments, which will cover a broad redshift range and
have significantly reduced statistical measurement errors compared to current
surveys in any particular redshift range.  The Dark Energy
Spectroscopic Instrument (DESI)\footnote{http://desi.lbl.gov/} is a
new MOS currently under construction for the 4-meter Mayall telescope
on Kitt Peak. DESI will be able to obtain 5000 simultaneous spectra,
which coupled with the increased collecting area of the telescope
compared with the 2.5-meter Sloan telescope, means that it can create
a spectroscopic survey of galaxies $\sim$ 20 times more quickly than
eBOSS. In 2020 the European Space Agency will launch the
Euclid\footnote{http://sci.esa.int/euclid} satellite mission. Euclid
is an ESA medium class astronomy and astrophysics space mission, and
will undertake a galaxy redshift survey over the redshift range
$0.9<z<1.8$, while simultaneously performing an imaging survey in both
visible and near-infrared bands. The complete survey will provide
hundreds of thousands images and several tens of Petabytes of
data. About 10 billion sources will be observed by Euclid out of which
several tens of million galaxy redshifts will be measured and used to
make galaxy clustering measurements.

In the current work we test the redshift weighting approach by
analysing a set of $1000$ mocks catalogues \citep{2015albert} designed
to match the eBOSS quasar sample. This quasar sample has a low density 
($82.6 \;\rm objects \rm/deg^{2}$) compared to that of recent galaxy samples, and
covers a total area over 7500 $\rm deg^2$. The quasars are highly biased
targets and we expect their bias to evolve with redshift,
$b(z) \propto c_1 + c_2( 1+ z) ^2 $, with constant values
$c_1 = 0.607 \pm 0.257 $, $c_2 =0.274 \pm 0.035$, as measured in
\cite{2017laurent}.
  
Although the mocks are not drawn from N-body simulations, they have
been calibrated to match one of the BigMultiDark (BigMD)
 \citep{2016kiplin}, a high resolution N-body
simulation, with $3840^3$ particles coverig a volume of
$(2500 h^{-1} Mpc)^3$. The BigMD simulations were performed using
GADGET-2 \citep{2005springel}, with $\Lambda$CDM Planck cosmological
constraints as a fiducial cosmology.
$\Omega_m =0.307$, $\Omega_b = 0.048206$, $\sigma_8 = 0.8288$, $ n_s = 0.96
$, $H_0 = 100 h \rm km s^{-1} \rm  Mpc^{-1}$ and $h = 0.6777$.  In
\cite{2015albert} the authors showed that EZ-mocks are nearly
indistinguishable from the full N-body solutions: they reproduce the
power spectrum within $1\% $, up to $k = 0.65 h \rm Mpc^{-1}$.  The
mocks are created using a new efficient methodology based on the
effective Zeldovich approximation approach including stochastic
scale-dependent, non-local and nonlinear biasing contribution.
The EZ mocks used for
the current analysis are the light-cone catalogues, realized on 7
different snapshots at $z= 0.9, 1.1, 1.3, 1.5 , 1.6 , 1.7 ,2.0 $.  The
full simulations incorporate the redshift evolution for $f$,
$\sigma_8$, the BAO damping and the non-linear density and velocity
effects.
  
In a companion paper \citep{2017Mio} we will apply the weighting scheme
to measure redshift-space distortions from the eBOSS DR 14 quasar data. In this
paper,
%
%
we validate the procedure and
test for optimality. By fitting to the evolution with a model for bias
and cosmology, we are able to fit simultaneously the
evolution of the growth rate $f(z)$, the amplitude of the dark matter density fluctuations $\sigma_8(z)$ and the galaxy bias
$b(z)$; breaking part of the degeneracy inherent in standard measurements of
$f\sigma_8$ and $b\sigma_8$ when only one effective redshift is
considered.  
  We show that the redshift weighting scheme gives unbiased
measurements.

The weights can be applied in both configuration or Fourier space. In
this paper, we focus  Fourier space, as there is some
evidence that this provides stronger redshift space distortions (RSD) constrains, given the
current scale limits within which the clustering can be modelled to a
reasonable accuracy \citep{Alam2016}.  In addition, the calculation of
the power spectrum moments is significantly faster than the
correlation function
\citep{bianchi2015, Scoccimarro2015}. Working in Fourier-space
requires a reformulation of the window selection to account for an evolving power
spectrum.

The paper is organized as follows; Section~\ref{optw} reviews the
derivation of optimal weights, presenting two schemes that differ in
the cosmological model to be tested. In Section~\ref{onez} we review
the redshift space power spectrum model at a single redshift; In
Section~\ref{mulze} we model the power spectrum and the window function to
obtain the redshift evolving power spectrum. In Section~\ref{fit} we
present the result of our analysis. 

\section{Optimal Weights}\label{optw} 

We make use of two different sets of weights; the first explores
deviations from the $\Lambda$CDM model by altering the evolution of
$\Omega_m$ in redshift. This model ties together growth and geometry,
but can also be used after fixing the expansion rate to
match the prediction of the $\Lambda$CDM model. The second
parametrizes the $f \sigma_8$ parameter combination measured by RSD,
allowing for a more standard test of deviations from
$\Lambda$CDM. Here, the growth and geometry are artificially kept separate as
$f \sigma_8$ only affects cosmological growth. 

The weight functions act as a smooth window on the data and allow us
to combine the information coming from the whole volume sampled. 
These weights are derived by minimising the error on the redshift
space distortion measurements, as predicted by a simple Fisher matrix analysis 
\citep{2016Mio}. 
Their derivation allows for the evolution with redshift of the cosmological parameters
we want to estimate from the data. Optimizing the measurements of the
parameters $\theta_i$ from the power spectrum moments $P_j$, we obtain the following weights,
\begin{equation}
w(z) = C^{-1} \partial P_{j}(z) /\partial \theta_i \label{defweight}.
\end{equation}
We assume the covariance matrix of $P$, $C$ to be parameter independent
and, in absence of a survey window, to be described as
\begin{equation} 
  C \sim (P_{\rm fid}  + 1/\overline{n} )^2 1/dV , 
\end{equation} 
for each volume element, $dV$ within the
survey. 
The weights can be seen as an extension of the FKP weights
presented in \cite{1994FKP}, which have the form,
\begin{equation}
w_{\rm FKP}(r) = \frac{1}{1+ \overline{n}(r) P(k)}, 
\end{equation}
by including the redshift component
$ \partial P_{j}(z) /\partial \theta_i $.

Note that the weights as they are reported in Eq.~(\ref{defweight}),
aim to compress different \textit{measurements} of the power spectrum
across a range of different redshifts. In fact, we apply weights
to each galaxy in order to avoid binning, by assuming
$w_{gal} = \sqrt{w_P}$, which relies on the scale-dependence of the
weights being smooth on the scale of interests for clustering.

\subsection{Optimal Weights for $\Omega_m$} \label{sec:omm}

As described in \citet{2016Mio}, it is empirically convenient to test
for deviations from the $\Lambda$CDM model by considering the
evolution of the matter density with redshift. To do this, we consider
a Taylor expansion up to second order about the fiducial model,
\begin{equation}
\dfrac{\Omega_m(z)}{ \Omega_{m,\rm fid}(z)}=  q_0 \left[ 1 + q_1y(z)  +\frac{1}{2} q_2 y(z)^2 \right],
\label{espomega}
\end{equation}
where $z_p$ is the pivot redshift and
$y(z)+ 1 \equiv \Omega_{m,\rm fid }(z) /\Omega_{m,\rm fid}(z_p) $. The
$q_i$ parameters correspond to the first and second derivatives of
$\Omega_{m}(z)|_{z_p}$, evaluated at $z_p$, and incorporate 
potential deviations about the fiducial model $\Omega_{m,\rm fid}$.

The choice of parameterising $\Omega_m$ (and hence the Hubble
parameter, the angular diameter distance and the growth rate) in terms
of $q_0$, $q_1$ and $q_2$ allows us to simultaneously investigate
small deviations using a common framework; e.g. departures from a
fiducial cosmology and geometry are accounted through the fiducial Hubble constant and
angular diameter distance  $H(\Omega_m)$, $D_A(\Omega_m)$; 
further,  modified gravity models can
be accounted through the growth rate, $f (\Omega_m)$.

By matching to the standard Friedman equation, we parametrize the
redshift evolution of the Hubble parameter in term of $\Omega_m(z)$
as,
\begin{equation}
H_{\rm}^2(z) = H^2_{0} \frac{ \Omega_{m,0 }(1+ z)^3 }{ \Omega_{m}(z) }.
\end{equation}
Assuming a flat Universe ($\Omega_{k} = 0 $) in agreement with CMB measurements \citep{2016Planck}, we have
$\Omega_{\Lambda}(z) = 1- \Omega_{m}(z)$. The subscript ``$0$''
denotes quantities evaluated at $z= 0$. For simplicity of notation we
omit the $q_i$ dependence from all the parameters: we refer to
$\Omega_m(z, q_i)$ as $\Omega_m(z)$, and we denote with
$\Omega_{m,\rm fid}$ the fiducial $\Lambda \rm CDM$ matter density.

For the scenarios considered, we assume the solution for the linear
growth factor $D(z)$ and the dimensionless linear growth rate $f$ have
the same dependence on $\Omega_{m}(z)$ as in the $\Lambda \rm CDM$ model,
\begin{equation} \label{eq:gf}
  g(z) \equiv ( 1+z)  D(z)  = \frac{5\Omega_{m}(z)H^3(z)}{2(1+z)^2} \int_z^\infty\mathrm{d}z'\frac{ (1+z')}{H^3(z')} 
\end{equation}  
\begin{equation}  \label{eq:ff}
 f(z) = -1 -\frac{ \Omega_{m}(z)}{2} +\Omega_{\Lambda}(z) + \frac{5\Omega_{m}(z)}{2g(z)} .
\end{equation} 


\begin{figure}{}
\includegraphics[scale=0.43]{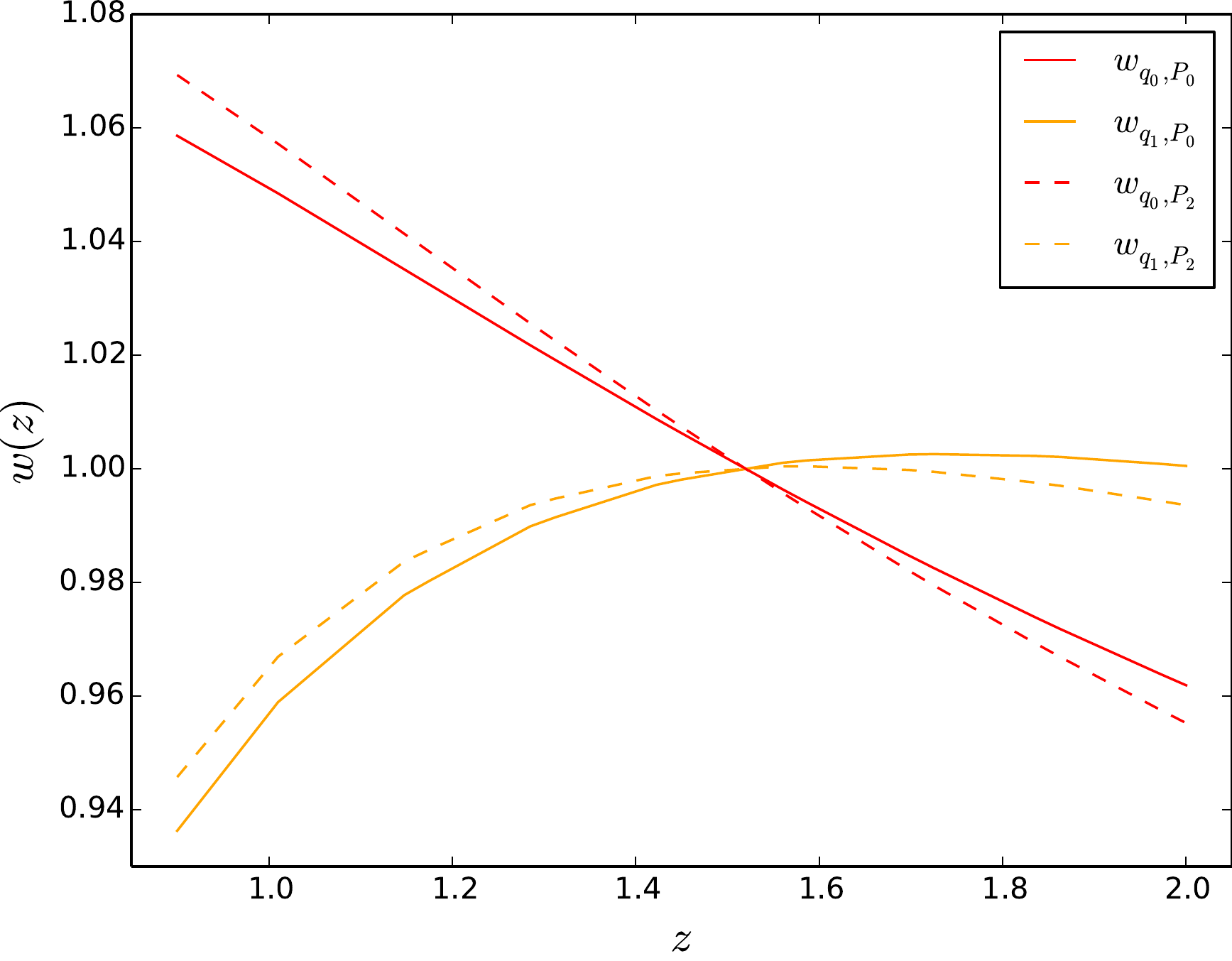}
\caption{The weights for the monopole  and quadrupole
 with respect to the $q_i$ parameters. 
  $\qquad \qquad $ $\qquad \qquad$ $\qquad \qquad$ }
 \label{mplot}
\end{figure}
Fig.~\ref{mplot} shows an example of the weights as derived in \cite{2016Mio},  that optimize the measurements of the $q_i$ parameters
in a $\Lambda$CDM fiducial background for a redshift-space power
spectrum. Since each multipole contains information about
$\Omega_m(z)$, our set of weights is derived to be optimal for the
first two non-null moments of the power spectrum on the Legendre
polynomial basis for each
$q_i$ parameter. Continuous lines indicate the weights for the monopole with respect to $q_0$ (red line) and $q_1$ (orange line); dashed lines indicate the weights for the quadrupole with respect $q_0$, $q_1$ (red and orange lines). 
All the weights are normalized to be equal 1 at the pivot redshift; 

\subsection{Optimal Weights for $f \sigma_8$} \label{fsig8}
%

RSD measurements constrain the amplitude of the velocity power
spectrum, and its cosmological dependence in the linear regime
is commonly parameterized by the product of the two parameters $f$ and
$\sigma_8$, which provides a good discriminator of modified gravity
models \citep{2009songperc}. We compare results obtained from the
$\Omega_m$ parametrisation with those derived using a set of weights
parametrised with respect to $[f \sigma_8] (z)$. In an analogous way
to the consideration in Section~\ref{sec:omm}, we can expand
$[f \sigma_8 ](z)$ about a fiducial model, so Eq.~(\ref{espomega}) becomes
\begin{equation}
  [f\sigma_8](z) = [f\sigma_8]_{fid}(z) p_0 \bigg( 1 + p_1 x + p_2 \frac{x^2}{2} \bigg)\label{fsev},
\end{equation}
where $x \equiv [f\sigma_8]_{fid}(z)/[f\sigma_8]_{fid}(z_p) -1 $.
The
$p_i$ parameters correspond to the first and second derivatives of
$ [f\sigma_8](z)|_{z_p}$, evaluated at $z_p$, and incorporate 
potential deviations about the fiducial model $[f\sigma_8]_{\rm fid}$.

We assume that in a $\Lambda$CDM model, $[f\sigma_8]_{\rm fid}$
behaves as 
\begin{equation}\begin{split}
  [f \sigma_8]_{\rm fid}(z) &= \bigg[-1 -\frac{ \Omega_{m, \rm
      fid}(z)}{2} +\Omega_{\Lambda, \rm fid}(z) 
    + \frac{5\Omega_{m, \rm fid}(z)}{2g_{\rm fid}(z)} \bigg]  \\
  &  \times \sigma_{8,0}\frac{ g_{\rm fid}(z)}{(1+z)^2},
\end{split}\end{equation}
with $g_{\rm fid}$, fiducial growth factor, 

\begin{equation}\begin{split}  \label{eq:gf}
g_{\rm fid}(z)
 =
\frac{5\Omega_{m,\rm fid}(z)H_{\rm fid}^3(z)}{2(1+z)^2} \int_z^\infty\mathrm{d}z'\frac{ (1+z')}{H_{\rm fid}^3(z')} 
\end{split}
\end{equation}

The galaxy bias parameter is assumed to be independent of $f$ and
$\sigma_8$. For simplicity, we consider $[b \sigma_8] $ to be
\textit{independent} from $[f\sigma_8]$ as well. 
 Considering e.g. the galaxy monopole  with respect to the linear matter power spectrum  $P$,
\begin{equation}
P_0 = \bigg(  [b \sigma_8]^2 + \frac{2}{3} [b \sigma_8] [f\sigma_8](z) +\frac{1}{5} [f\sigma_8]^2(z) \bigg)P(k),
\end{equation}
the dependence on the $p_i$ parameters is given only through $[f\sigma_8]$.  
We derive the set of weights  by taking the derivative of $P_0$, $P_2$, $P_4$ with respect to $p_1$, $p_2$, $p_3$. 
For completeness we include the weights here as they were not included in \cite{2016Mio}. 
\begin{equation}\begin{split}
w_{i, q_0 } =  N _i, \;\; w_{i, q_1 } = N_i y,  \;\; w_{i, q_2 } = N_i \frac{1}{2} y^2,
\end{split}
\end{equation}
where
\begin{equation}
  N_0 \equiv \left( \frac{2}{3} [b \sigma_8 ]  + \frac{2}{5} [f\sigma_8](z) \right) [f\sigma_8](z),
\end{equation}
\begin{equation}
  N_2 \equiv \left( \frac{4}{3}[ b \sigma_8]   + \frac{8}{7} [f\sigma_8](z) \right) [f\sigma_8](z),
\end{equation}
\begin{equation}
  N_2 \equiv \left( \frac{16}{35} [f\sigma_8](z) \right) [f\sigma_8](z).
\end{equation}
%


A strong effect on the set of weights is caused by  the
assumptions we make for galaxy bias. If we set the bias as an unknown parameter,
and we marginalize over it, then we cannot  deduce any information
about structure growth from the amplitude of the density power
spectrum. This is the case for the expansion around $[f\sigma_8]$,
where we considered $[b \sigma_8]$ and $[f \sigma_8]$ as independent
parameters. However, if we constrain $b(z)$ to match a fiducial model,
we will derive weights that make use of the information coming from
the amplitude of the power spectrum. For the expansion around
$\Omega_m$, we can choose  whether or not to include this
information.

\section{Modelling the anisotropic galaxy power spectrum at a single redshift}\label{onez}

We model the power spectrum using perturbation theory (PT) up to
1-loop order.  We include the non linear redshift space distortions
effects as in \cite{2004scocci} and \cite{2010TSN} (TSN model),
\begin{equation}
\begin{split}
P_{\rm g}(k,\mu) &= \exp\left\{-(fk\mu\sigma_v)^2\right\}\big[P_{{\rm g},\delta\delta}(k)\\
&\;\;\;\; + 2f\mu^2P_{{\rm g},\delta\theta}(k) + f^2\mu^4P_{\theta\theta}(k)\\
&\;\;\;\; + b^3A(k,\mu,\beta) + b^4B(k,\mu,\beta)\big], 
\label{eq:taruya}
\end{split}
\end{equation}
where $\mu$ is the cosine of the angle between the wave-vector $\textbf{k}$ and
the line of sight.  $P_{\theta \theta } $ and $P_{\delta \theta }$ are
the \textit{velocity-velocity} and \textit{matter-velocity} power
spectra terms that correspond to the extended linear model of 
\cite{kaiser1987} as derived in \cite{2004scocci}.  $\theta$ denotes
the Fourier transform of the comoving velocity field divergence,
$\theta(\textbf{k}) \equiv -i \textbf{k} \cdot \textbf{u}(k) $ where
$\nabla \textbf{u } = -\nabla \textbf{v }/\big[a f(a) H(a) \big] $ with
velocity field  $v$ and dimensionless linear growth rate $f$.  The
exponential term represents the damping due to the ``Fingers of God''
effect, where $\sigma_v$ denotes the velocity dispersion term, here
treated as free parameter.  The $A$, $B$ terms come from the TNS model
which take into account further corrections due to the non linear
coupling between the density and velocity fields \citep{2010TSN}.
Note that at linear level
$P_{\theta \theta } = P_{\delta \theta } = P_{\delta \delta }$.

We model $P_{{\rm g},\delta\delta}$ and $P_{{\rm g},\delta\theta}$ as 
\begin{equation}
\begin{split}
P_{{\rm g},\delta\delta}(k) &= b^2P_{\delta\delta}(k) + 2b_2b P_{b2,\delta}(k) 
+ 2b_{s2}b P_{bs2,\delta}(k)\\
& + 2b_{\rm 3nl}b \sigma_3^2(k)P(k) + b^2_2P_{b22}(k)\\
& + 2b_2b_{s2}P_{b2s2}(k) + b^2_{s2}P_{bs22}(k) + S,
\label{eq:40}
\end{split}\\
\end{equation}

\begin{equation}
\begin{split}
P_{{\rm g},\delta\theta}(k) &= b P_{\delta\theta}(k) + b_2P_{b2,\theta}(k) 
+ b_{s2}P_{bs2,\theta}(k)\\
& + b_{\rm 3nl}\sigma_3^2(k)P (k),
\label{eq:41}
\end{split}
\end{equation}
%
The bias is modelled following recent studies
\cite{2012bias, 2012bald} that showed the importance of non-local
contributions. We account for those effects introducing as galaxy bias
parameters: the linear $b$, second order local $b_2$, non local
$b_{s2}$, and the third order non-local $b_{\rm 3nl}$ bias parameters,
and the constant stochasticity shotnoise term $S$. We numerically evaluate the
non-linear matter power spectra, $P_{\delta\delta}$,
$P_{\delta\theta}$, $P_{\theta\theta}$, at $1$-loop order in standard
perturbation theory (SPT) using the linear power spectrum input from
CAMB \citep{Lewis2002ah}.

In the current analysis we make use of the first three non-zero
moments of the power spectrum, projected into an orthonormal basis of
Legendre polynomials $\mathcal{L}_\ell (\mu)$ such that,
\begin{equation}
P_\ell (k)= \frac{2 \ell +1 }{ 2 } \int_{-1}^{1} d \mu P(k, \mu) \mathcal{L}_\ell (\mu),
\label{multipoles}
\end{equation}
with the monopole $\ell = 0$, quadrupole $\ell = 2$ and hexadecapole
$\ell = 4$, respectively.
In this paper we do not consider geometrical deviations and we are only concerned with growth measurements in a fixed background. However, we note that such deviations can be included as follows. 
The geometrical deviations from the fiducial cosmology can be included through the Alcock-Paczynski effect, \citep{1979AP}. Here, revised
mode numbers $k'$, $\mu'$ for the cosmological model being tested, are
related to those observed $k$, $\mu$ assuming the fiducial cosmology
by the transformations
\begin{equation}\begin{split}
k'     &= \frac{k}{\alpha_\perp} \big[ 1 + \mu^2 \bigg( \frac{\alpha_\perp^2}{\alpha_\parallel^2 } -1 \bigg)    \bigg]^{1/2} \\
\mu' &= \frac{\mu \alpha_\perp}{\alpha_\parallel} \bigg[ 1 + \mu^2 \bigg(  \frac{\alpha_\perp^2}{\alpha_\parallel^2 } -1 \bigg)    \bigg]^{-1/2} 
\end{split}
\end{equation}
where the scaling factors   $\alpha_\parallel $ and $\alpha_\perp$ are defined as
\begin{equation}  \label{eq:alpha}
\begin{split}
 \alpha_\parallel  &= \frac{H^{\rm fid} (z)}{H(z)}, \\
 \alpha_\perp      &= \frac{D_A(z) }{D^{\rm fid}_A(z)}.
\end{split}
\end{equation}
 
By applying the transformations of Eq.~(\ref{eq:alpha}) to
Eq.~(\ref{multipoles}), the multipoles at the observed $k$ and $\mu$, relate to the power spectrum at the true
variables $k'$ and $\mu'$ through
\begin{equation}
P_\ell(k) = \frac{(2 \ell + 1 )}{2 \alpha^2_\perp \alpha_\parallel} \int_{-1}^{1} \;d\mu\, P_g (k', \mu' ) \mathcal{L}_\ell (\mu).
\label{multip2}
\end{equation}

\section{Modelling the evolving galaxy power spectrum}\label{mulze}

\subsection{Redshift weighted multipoles without window function}\label{mulmeas}

We model the redshift dependence of $f,$ $\sigma_8$,
$\alpha_\parallel$, and $\alpha_\perp$ as described above, and the bias evolution (see Sec. \ref{biasevol}).
 In principle we can compute the weighted multipoles by integrating the power spectrum moments as given
in Eq.~(\ref{multip2}) over redshift, including the redshift
weighting,
\begin{equation}
P_{\ell w_{\ell, q_j} }  =\int d\,z \; P_\ell (k  , z) w_{\ell, q_j}.
\end{equation}

However, by considering the power spectrum as a redshift evolving
quantity we need to redefine the survey window function. It is easy to
demonstrate that, by introducing an evolving power spectrum
$P(k, z)$ into equation 2.1.4 of \cite{1994FKP}, is no
longer possible to derive the convolution relation 2.1.6 between the
survey window function and the true power spectrum.  We now consider
how to allow for a generalized window function for redshift evolving
power spectrum multipoles.

\subsection{Redshift weighted multipoles including the survey window effect}\label{window}

We study the window function for the evolving power spectrum using a
generalized Hankel transformation between power spectrum and
correlation function moments, where the window applied is also
decomposed into a set of multipoles.  This is an extension of the work
by \cite{2017wilson} and \cite{2017beutl}, which presented a method to
convolve model power spectra with the window function for a
non-evolving power spectrum. We consider the case where the underlying
correlation function $\xi$ is dependendent on both the separation
$r = |r_i - r_j|$ (with $r_i$ and $r_j$ position of galaxies of each
pair) and the mean redshift of each galaxy pair
$\xi[r_i(z_i), |r_i - r_j|]$. Here we have assumed that cosmological
evolution is negligible over the range of redshifts covered by every
pair, so we can quantify the clustering of each using the correlation
function at the mean redshift.

The multipole moments of the power spectrum in the local
plane-parallel approximation can be written as,
\begin{eqnarray}
	\begin{split}
		\hat{P}_\ell(k) =&\frac{2\ell+1}{2} \int d\mu_k\int \frac{d\phi}{2\pi} \int d\vc{x}_1 \int d\vc{x}_2e^{i\vc{k}\cdot \vc{x}_1}e^{-i\vc{k}\cdot \vc{x}_2}\\
		&\llan \delta(\vc{x}_1) \delta(\vc{x}_2)W(\vc{x}_1)W(\vc{x}_2)\rran    \mathcal{L}_{\ell}(\hat{\vc{k}}\cdot\hat{\vc{x}}_h)  \\
                =&\frac{2\ell+1}{2} \int d\mu_k\int \frac{d\phi}{2\pi}\int d\vc{x}_1 \int d\vc{s} \;\times\\
		&\left(\sum_L\xi_L[s, z(x_1)] \mathcal{L}_{L}(\hat{\vc{x}}_h\cdot\hat{\vc{s}})  \right) \;\times \\
		& W(\vc{x}_1)W(\vc{x}_1+\vc{s}) e^{-i\vc{k}\cdot \vc{s}} \mathcal{L}_{\ell}(\hat{\vc{k}}\cdot \hat{\vc{x}}_h),
                \label{mull}
              \end{split}
\end{eqnarray}
where $\int d\mu_k$ is the integral over all the possible cosine angles between
$\hat{\vc{k}}$ and $\hat{\vc{x}}_h$ and $W$ defines the
mask. $\xi_L$ denotes the correlation function moments in the Legendre basis. 
Note that Eq.~(\ref{mull}) differs from
equation A.16 in \cite{2017beutl}, only in the $\xi_L[s, z(x_1 )] $
term; for a single redshift slice we would only have $\xi_L(s)$. We
make use of the relations,
\begin{eqnarray}
e^{-i \vc{k}\cdot \vc{s}} = \sum_p (i)^p (2p+1)j_p(ks)\mathcal{L}_{p}(\hat{\vc{k}}\cdot \hat{\vc{s}}),
\label{ap1}
\end{eqnarray}
\begin{eqnarray}
\int d\mu_k\int \frac{d\phi}{2\pi} \mathcal{L}_{\ell}( \hat{\vc{k}}\cdot\hat{\vc{x}}_h ) \mathcal{L}_{p}(\hat{\vc{k}}\cdot \hat{\vc{s}}) = \frac{2}{2\ell+1}\mathcal{L}_{\ell}(\hat{\vc{s}}\cdot\hat{\vc{x}}_h)\delta_{\ell p},
\label{ap2}
\end{eqnarray}
which, when combined with  Eq.~(\ref{mull}), give
\begin{equation}\begin{split}
\hat{P}_\ell(k)  & = i^\ell (2\ell+1) \int d\vc{x}_1 \int d\vc{s} \; \sum_L \xi_L[s, z(x_1)]    j_\ell(ks) \; \times \\
		      &W(\vc{x}_1)W(\vc{x}_1+\vc{s})  \mathcal{L}_{\ell} (\hat{\vc{x}}_h \cdot \hat{\vc{s}})  \mathcal{L}_{L} (\hat{\vc{x} }_h\cdot\hat{ \vc{s} } ).
\label{peelw}		      
\end{split}\end{equation}
Substituting the Bailey relation,
$\mathcal{L}_{\ell}\mathcal{L}_{p} = \sum_t a^{\ell}_{p
  t}\mathcal{L}_{t}$, Eq.~(\ref{peelw}) becomes,
\begin{equation}\begin{split}
\hat{P}_\ell(k)  =& i^\ell (2\ell+1) \int d\vc{x}_1 \int d\vc{s} \;  \sum_L \xi_L[s, z(x_1)]   j_\ell(ks)  \; \times \\
		        &W(\vc{x}_1)W(\vc{x}_1+\vc{s})  \mathcal{L}_{\ell} (\hat{\vc{x}}_h \cdot \hat{\vc{s}}) \sum_t a^{\ell}_{L t}\mathcal{L}_{t}(\hat{\vc{x} }_h\cdot\hat{ \vc{s} } ).\\
		      =&i^\ell (2\ell+1) \int 2\pi s^2 ds\;  j_\ell(ks) \sum_L \sum_t a^{\ell}_{L t} \int d\mu_s \int \frac{d\phi}{2\pi} \\
		        &\times  \int d\vc{x}_1\; \xi_L[s, z(x_1)] W(\vc{x}_1)W(\vc{x}_1+\vc{s})\mathcal{L}_{t}(\hat{\vc{x} }_h\cdot\hat{ \vc{s} } )
\end{split}\end{equation}

At this stage, in contrast to Eq. $A.19$ in \cite{2017beutl}, we
cannot bring $\xi_L$ out of the integral over $x_1$. Since we are not
able to decouple the mask from $\xi$, in principle, we would have to
compute the integral over $x_1$ for every model $\xi$ fitted to the
data. However we can reduce drastically the computational time
required by assuming that $\xi$ is well behaved such that we can split
the integral over $x_1$ into a sum over a small number of $x_i$
ranges. This is different from measuring the clustering in shells - we
are still calculating and modelling the power spectrum as a
continuously weighted function calculated using every galaxy pair; we
are simply making an assumption about the smooth behaviour in redshift of the expected
clustering.
\begin{equation}\begin{split}
\hat{P}_\ell(k)  
=&i^\ell (2\ell+1) \int 2\pi s^2 ds\;  j_\ell(ks) \sum_L \sum_t a^{\ell}_{L t} \int d\mu_s \int \frac{d\phi}{2\pi} \\
		        &\times  
		     \sum_{i}   \int_{x_i} d\vc{x}_1\; \xi_L(s, z(x_i)) W(\vc{x}_i)W(\vc{x}_i+\vc{s})\mathcal{L}_{t}(\hat{\vc{x} }_h\cdot\hat{ \vc{s} } ).
\end{split}\end{equation}
Assuming that $\xi_L(s, z(x_i))$ is constant over each sub-integral
range $x_i $ we can take it out of the integrals,
\begin{equation}\begin{split}
&\hat{P}_\ell(k)  
=i^\ell (2\ell+1) \int 2\pi s^2 ds\;  j_\ell(ks) \sum_L \sum_t a^{\ell}_{L t} \times \\
       &    \sum_{i}  \xi_L(s, z(x_i) )  \int d\mu_s \int \frac{d\phi}{2\pi} 
		   \int_{x_i} d\vc{x}_i\;  W(\vc{x}_i)W(\vc{x}_i+\vc{s})\mathcal{L}_{t}(\hat{\vc{x} }_h\cdot\hat{ \vc{s} } ),
		   \label{wiwi}
\end{split}\end{equation}
and redefine the sub-window function multipoles $W_{p, z_i}^2(s)$ for
$p = 0,2, 4.. $ as 
\begin{equation}
  \begin{split}\label{e30}
    W_{p,z_i }^2(s)  =&\frac{2p+1}{2} \int d\mu_s  \int \frac{d\phi}{2\pi}\int_{z_i} d\vc{x}_i \\
    & \times W(\vc{x}_i)W(\vc{x}_i+\vc{s})\mathcal{L}_{p}(\mu_s).\\
  \end{split}
\end{equation}
Using the definition of the sub-window function multipoles of
Eq. \ref{e30}, we can write Eq.~(\ref{wiwi}) to be
\begin{equation}
  \begin{split}
    \hat{P}_\ell(k) &= i^\ell (2\ell+1)\int 2\pi s^2ds\;  j_\ell(ks)  \;\times\\
    &\;\;\;\;\sum_L\sum_t \frac{2}{2t+1} a^{\ell}_{L t}   \sum_{i}  \xi_L(s, z_i)
    W_{t , z_i}^2(s).
  \end{split}
\end{equation}
which generalizes Eq. A.23 in \cite{2017beutl} to the case of a
redshift-evolving power spectrum.

 

In this study we make use of the redshift weights to compute the
weighted multipoles $\hat{P}_{\ell, w_j} $: we include the weights in
the model by inserting them in the sub-window multipoles. As these are
calculated using a pair-counting approach we apply the weights to the
pairs as for the data. The binning of the mask does not affect our
ability to use continuous redshift weighting on the data and in the
model. Thus, we do not bin our redshift weights. However, we do assume
that the weights are scale independent, so they can be applied to the
galaxies assuming $w_g = \sqrt{w_P}$.

Note that we include a fiducial model for the evolution of the power
spectrum when calculating the weights; however, even in the case that
the fiducial model does not match accurately the data, the weights
would not bias the measurements; instead we would have that the
weights we are using are not optimal.

\subsection{Bias evolution }\label{biasevol}
Differently from the other quantities  the galaxy bias, $b$,  requires a more careful study before including it in the evolving power spectrum: $b(z)$  strongly depends on the targets and there is not a cosmological constraint; for this reason we should allow for more freedom in the form that it can take. 
  
    In  \cite{2017Mio}  we compared the weights for different $b(z)$ relations and  showed that the weights are not significantly sensitive to the different  $b(z)$ considered; 

The fitting formula for the linear bias  parameter of the quasar sample  suggests that the linear bias redshift evolves as, \citep{2017laurent}, 
\begin{equation} 
b(z) = 0.53 + 0.29(1 + z )^2. 
\end{equation}
  We model the evolution of $b$ about the pivot redshift  times $\sigma_8$ as, 
 \begin{equation}
 b \sigma_8(z) =  b \sigma_8(z_{\rm p}) + \partial b \sigma_8 /\partial z |_{z_p} (z - z_{\rm p}) + ...   .
 \end{equation}
 We  neglect the redshift dependence for the non linear bias parameter $b_2$, so we assume this is constant with redshift, $b_2 \sigma_8(z_{\rm piv})$.  We fix the  2nd-order non-local bias, $bs_2$ and 3rd-order non-local bias, $b_{3nl}$  terms to their predicted  values according to non-local Lagrangian models,\citep{2012bias, 2012bald}, 
\begin{equation}\begin{split}\label{nlbias}
b_{s2} &= -\frac{4}{7} (b -1 ), \\
b_{3nl} &= \frac{32}{315} (b -1 ).
 \end{split}\end{equation}

%
%
\section{Fitting to the mock data}\label{fit}

\subsection{Power spectrum measurement} 

To compute the power spectrum moments with respect to the line of sight (LOS), we make
use of the estimator introduced in \cite{bianchi2015}.  This  fourier transform (FT) - based
algorithm uses multiple FTs to track the multipole moments, in the
local plane-parallel approximation where we have a single LOS for each
pair of galaxies. This estimator has been already used in recent
analysis \citep{2017beutl}, that confirmed the advantages of using
such decomposition: it reduces the computational time from $N^2$
associated to naive pair counting analysis \cite{yamamoto2006} to
$\sim N {\rm log} N$.  

Redshift weights are included in the estimator, by defining the
weighted galaxy number density as $n_g (\textbf{r}) w$.  As discussed
in Section~\ref{optw} we have derived the galaxy weights from the
square-root of the power spectrum weights, under the assumption that
the scale dependence in the weights is smooth compared to the scale of
interest for our clustering measurements.  

The result is a set of  weighted multipoles,
$P_{0,2, w_{0,1,2}}$, where each $P_{i, w_j}$ corresponds to a
particular set of weights that optimizes each of the $q_i$ or $p_i$
measurement,  i.e. for the set of weights $w_{i, q_j}$ (or
$w_{i, p_j} $ for the $f\sigma_8$ weights) functions and we build a
data vector $\Pi$ as,
\begin{equation}
 \Pi^T =( P_{0, w_0,q_0}, P_{0, w_0,q_1}, P_{0, w_0,q_2} \; ...  \; P_{2, w_2,q_2})^{\rm T}.
\end{equation}  

\subsection{Covariance matrix estimation}
\label{COV}
We evaluate the covariance matrix for the data vector $\Pi^T$ using
1000 EZ mock described in Section~\ref{intro}.  For each mock, we
compute the weighted monopole and quadrupole moments for each set of
optimal redshift weights, for $n_b =10$ $k$-bins in the range of
$k = 0.01 - 0.2 h \rm Mpc^{-1}$. From these, we derive the covariance
matrix as
\begin{equation}\begin{split}
C = \frac{1}{N_{\rm T} -1 } \sum_{n=1}^{N_{\rm T}} \big[  P_{n, \ell, w_{\ell, q_t} }(k_i) - \hat{P}_{ \ell, w_{\ell, q_t} }(k_i) \big] \\
 \times \big[  P_{n, \ell, w_{\ell, q_t} }(k_j) - \hat{P}_{ \ell, w_{\ell, q_t} }(k_j) \big],
\label{covmat}
\end{split}\end{equation}
where $N_T = 1000$ is the number of mock catalogues, $w_{\ell, q_t}$
denotes each set of weights for each parameter $q_t$ (or $p_t$) and
$\hat{P}_{ \ell, w_{\ell, q_t} }(k_i) = \frac{1}{N_{\rm T} } \sum_{n=1}^{N_{\rm T} } P_{ n, \ell, w_{\ell, q_t}}(k_i)$.

Note that when inverting the covariance matrix we include the Hartlap factor \citep{hartlap2007} 
to account for the fact that $C$ is inferred from mock catalogues. 

\subsection{Maximising the Likelihood}

Since each weighted multipole $P_{i, w_{i, q_j}}$ is optimized with
respect to a particular piece of information (e.g. $\Omega_m[z]$),we
jointly fit all three $q_i$ (or $p_i$) parameters simultaneously. We compare the
measured $\Pi^T$ to modelled weighted power spectra multipoles,
convolved with the window function as explained in
Section~\ref{window}.  We assume a Gaussian likelihood and minimize
\begin{equation}
  \chi^2 \propto ( \Pi -  \Pi_{\rm model} )^{\rm T} C^{-1} (\Pi -  \Pi_{\rm model} );
\end{equation}
Where $\Pi_{\rm model}$ refers to the window convolved
$P_{i, w_{i, q_j}}$.  The $C^{-1}$ term corresponds to the joint
covariance derived in Eq.~(\ref{covmat}). We repeat the fit for both the 
$\Omega_m$ and $f\sigma_8$ optimized sets of weights.

In the current analysis we limit ourselves to  linear order
deviations about our fiducial $\Lambda$CDM model, for both
$f\sigma_8(z)$ and $\Omega_m(z)$ described in Sections~\ref{sec:omm}
and~\ref{fsig8}, since the data cannot capture second-order
deviations. We discuss this further in Section~\ref{discuss}.
%
%
%
%
%
\section{Measuring RSD with the evolving galaxy power specturm}
\label{noap}

 The fits presented in this section are performed using a Monte Carlo Markov Chain (MCMC) code, implemented to efficiently account for the degeneracies between the  parameters;  in all the fit performed we select a range between $k = 0.01 - 0.2 h  \rm Mpc^{-1}$. For each scenario explored we run 10 independent chains, satisfying the Gelman-Rubin convergence criteria \citep{Gelman92} with the requirement of $R -1 < 10^4 $; where $R$ corresponds to the ratio between the variance of chain mean and the mean of chain variances.  
 All the results presented are obtained after  marginalizing on  the full set of parameters, including the nuisance parameters (shotnoise and velocity dispersion). All the contour plots are produced using the public getdist libraries\footnote{http://getdist.readthedocs.io/}. 
 
We fit the weighted monopole and quadrupole computed on a
subset of 20 EZ mocks, for both the $\Omega_m$ and the
$f\sigma_8$-optimized weights, while keeping the distance-redshift
relation fixed to the fiducial cosmology, i.e.
$\alpha_\parallel =\alpha_\perp = 1 $. 

We do not consider the full set of $1000$ EZ mocks for the following reasons; first we are limited  by the EZ-mock accuracy in describing non linearities in galaxy bias and  velocities;  further by the  accuracy in the light-cone describing  the redshift evolution  for  $f \sigma_8 $ which is included as a step function. Thus we do not believe that the mocks supports us looking at deviations from the model at better accuracy than this.
  However, the error on our constraints is  still $1/\sqrt{20}$ smaller than what we expect on the eBOSS quasars constraints.
The analyisis has been performed on different subset of 20 mocks out of the 1000 available to verify that the outcomes do not depend on a particular subsample choice.

%
 %
%


Our analysis is presented as follow;  
in \ref{aa} we present the result obtained with the $\Omega_m$weights fitting for  $ q_0$, $q_1$, $b \sigma_8 (z_{\rm p})$, $\partial b \sigma_8/\partial z|_{z_p} $, $b_2$, $\sigma_v$, and shotnoise $S$.  In parallel we present the fit for $p_0, p_1$, $b \sigma_8 (z_{\rm p})$, $\partial b \sigma_8/\partial z|_{z_p} $, $b_2$, $\sigma_v$  when applying the $f\sigma_8$ weights.

In \ref{bb}  we investigate  the impact of the  bias assumption on the contraints, showing a comparison between bias    evolving and constant with redshift. 

In \ref{cc} we compare the  results obtained with the redshift weights approach with the analysis performed considering one \textit{constant} redshift slice i.e. considering all the parameters ($f\sigma_8, b\sigma_8, \sigma_v, b_2, S$) in the power spectra  at their value at the pivot redshift  $z = 1.55$ and  applying FKP weights only (for simplicity of the notation from now on we refer to this  as \textit{traditional} analysis).

Differently from \cite{2016Zhu}, we compare the redshift weights analysis with the standard analysis used for previous RSD measurements (see e.g. \cite{2017beutl}) rather than testing the weights $w_{q,i}, w_{p,i } =1$. 
The main focus of this work is to test that our analysis is not biased by introducing evolution in the power spectrum and in the window function. 
We rely on the Fisher matrix theory correctly selecting the set of weights optimal with respect  to the $q_i, p_i$ errors.  
%


\subsection{Redshift weights fit}\label{aa}

\begin{figure}
\includegraphics[scale=0.4]{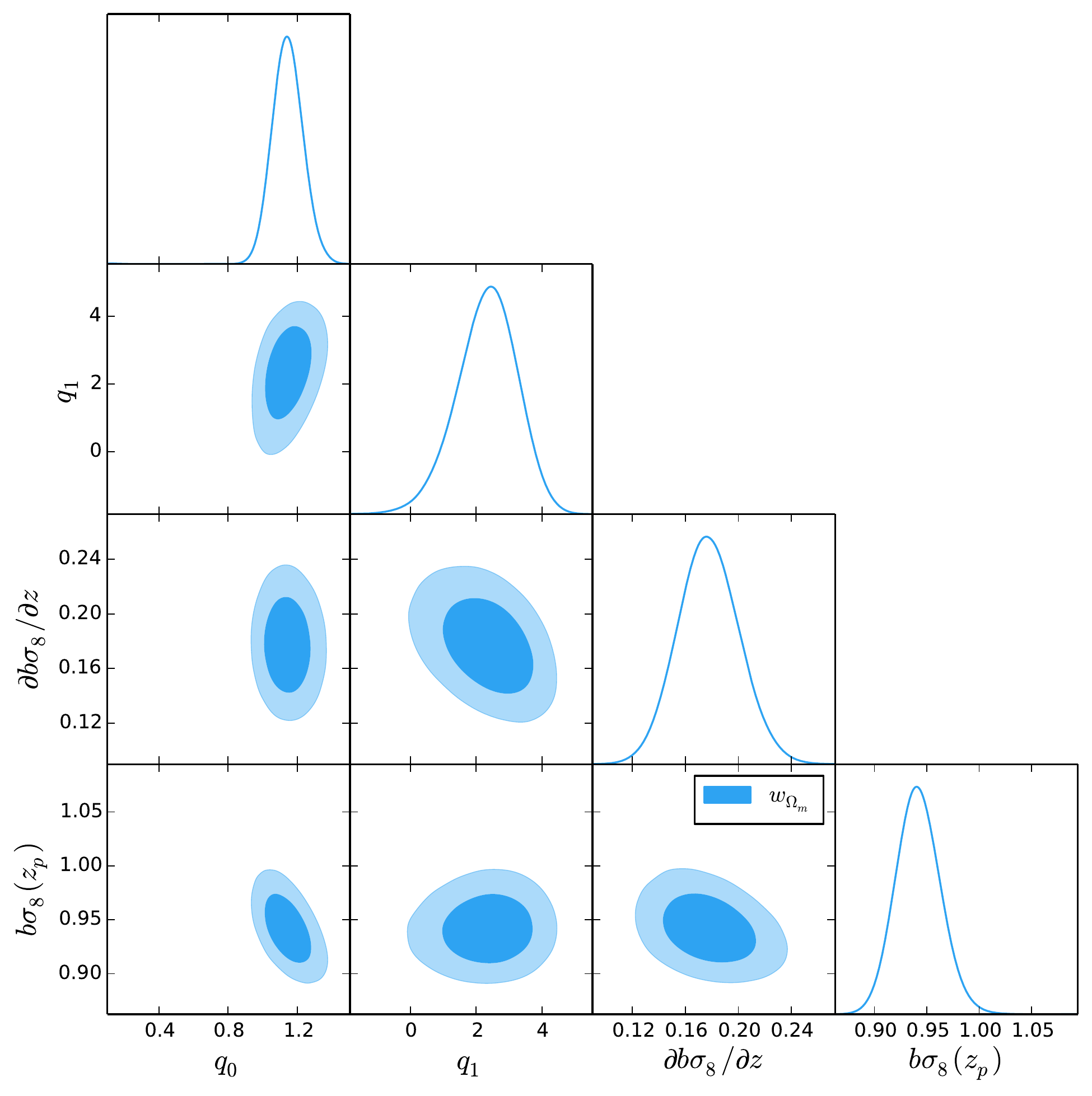}
\caption{ Likelihood distributions  for the analysis of the average of 20 EZ mock. We show the results for  $ q_0$, $q_1$, $b\sigma_8(z_{p})$, $\partial b\sigma_8/\partial z $, marginalized over the full set parameters (including $b_2 \sigma_8(z_p)$, $\sigma_v$, $S$ not dispayed here). We multi-fit two weigthed monopoles and two weigthed quadrupoles (one for each weight function ($w_{0,p_i}$, $w_{2,p_i}$) The fitting range is $ k = 0.01 - 0.2 h \rm Mpc^{-1}$ for both the monopole and quadrupole. 
}
\label{figura1}
\end{figure}

Fig. \ref{figura1} shows the posterior likelihood distributions  from the analysis performed with  the  set of redshift weights optimized to constrain   $\Omega_m(z)$ (blue contour plots), using the monopole and the quadrupole;  we fit for  $q_0$, $q_1$ which describe up to  linear order deviations in the evolution of $\Omega_m(z)$ according to $\Lambda \rm CDM$ model;   we also vary the galaxy bias parameters
  modelled as  in section \ref{biasevol}, while we fix the  2nd-order non-local bias, $bs_2$ and 3rd-order non-local bias, $b_{3nl}$  terms as shown in Eq. \ref{nlbias}. 
   To summarize we fit  for 7 parameters:  $q_0$, $q_1$, $b \sigma_8(z_{\rm p}) $, $\partial b \sigma_8 /\partial z|_{z_p}$, $b_2 \sigma_8(z_p)$, $\sigma_v$, and shotnoise $S$.  

%
%
 %
%
%
%
%
%
%
%
%
%
%
%
%
%
%
%
%

%
%
%
%
%
%
%
%
%
\begin{figure}
\includegraphics[scale=0.4]{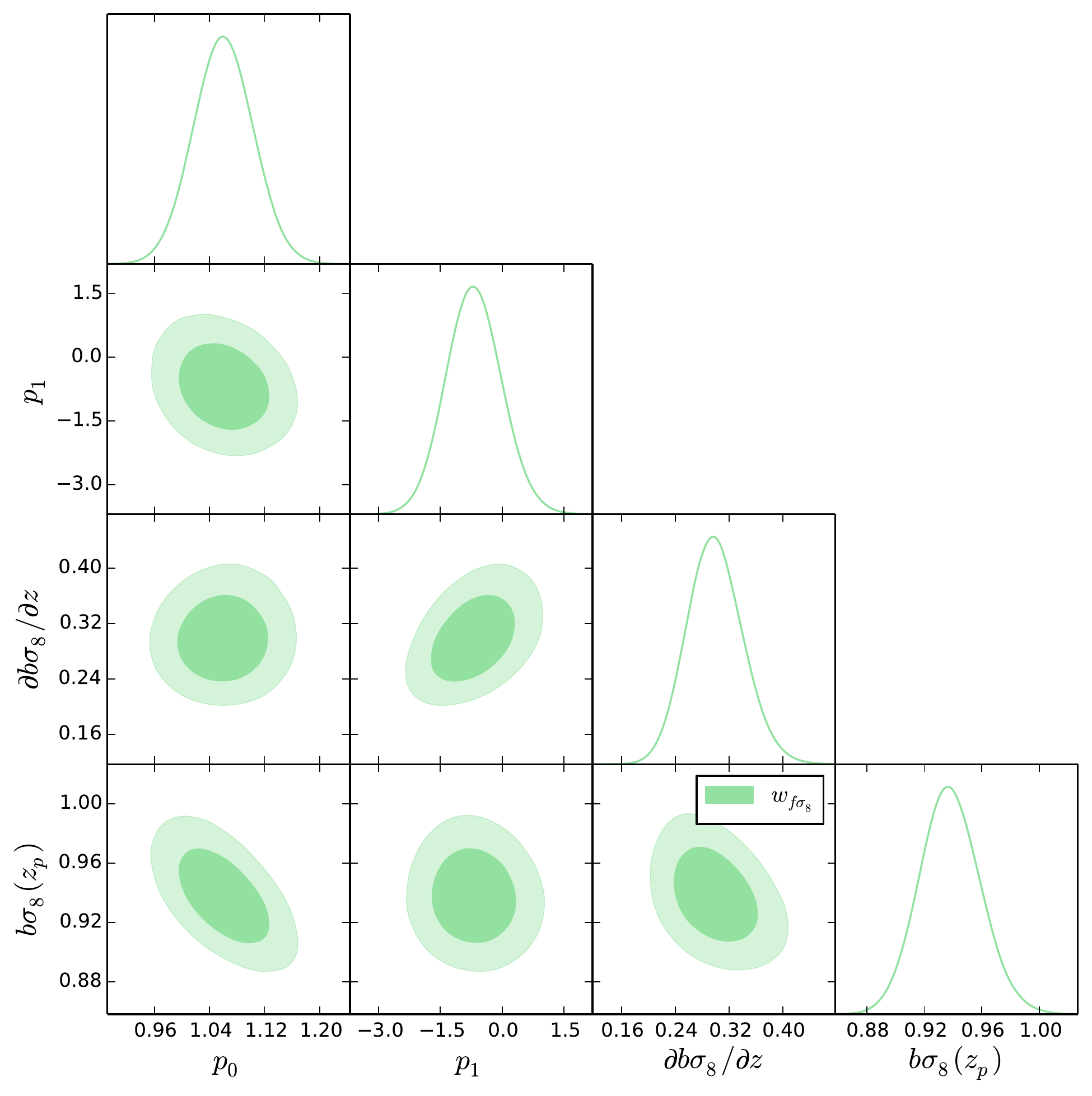}
\caption{Likelihood distributions  for the analysis of the average of 20 EZ mock. We show the results for  $ p_0$, $p_1$, $b \sigma_8 (z_{p})$, $\partial b\sigma_8/\partial z$, 
 marginalized on the full set parameters (including $b_2 \sigma_8(z_p)$, $\sigma_v$, $S$ not dispayed here). We multi-fit two weigthed monopoles and two weighted quadrupoles (one for each weight function ($w_{0,p_i}$, $w_{2,p_i}$) The fitting range is $ k = 0.01 - 0.2 h \rm Mpc^{-1}$ for both the monopole and quadrupole. 
}
\label{figura2}
\end{figure}

 Fig \ref{figura2} presents the results of the analysis while using the  set of redshift weights optimized to constrain $f\sigma_8(z)$, as  introduced in Sec.  \ref{fsig8}; the structure is the same as in  Fig. \ref{figura1}. 
 We fit for $p_0$, $p_1$ to constrain $f\sigma_8(z) $ deviations about the fiducial $f\sigma_8(z) $ according to $\Lambda \rm CDM$; we also fit  for  $ b\sigma_8 (z_{\rm p})$, $\partial b \sigma_8(z) /\partial z $, $b_2 \sigma_8(z_p)$, $\sigma_v$, $S$, 7 parameters in total as for the other set of weights.

The resulting posteriors in both  Figures \ref{figura1} and \ref{figura2}  show a correlation between the zero order parameters,
  $q_0$ ($p_0$) and  $b \sigma_8 (z_{\rm piv})$, of magnitude of $ \sim 0.5 $. We also detect a relevant anti-correlation  $ \sim - 0.4 $ between the slope parameter $q_1$  ($p_1$) and the gradient  $\partial b \sigma_8(z) /\partial z $.  We tested and confirmed that those correlations lead to a mild dependency between the assumed bias model (linear and non linear in $k$ and in $z$) and the slope parameter $q_1$ ($p_1$) without 
  however affecting (within $\sim 1 \sigma$) the constraints on $f\sigma_8$. In \ref{bb} we illustrate this in more details. 
  %
%
%
%
 %
 Due to the stepwise implementation of the growth rate and bias model in the mocks, the fiducial values of  $q_{0}, q_{1}$ ($p_{0}, p_1$) are not well defined.
Therefore, we do not display an expected value for $p_i$ and $q_i$ as those cannot be inferred from the $f\sigma_8$ evolution  included as a non-smooth step function in the mocks. However, within 1 to 2sigma we recover the smooth $\Lambda \rm CDM$ expectation values of $q_0=1$ and $q_1 = 0$.
%
%
 %
%
%
 \begin{figure}
\includegraphics[scale=0.4]{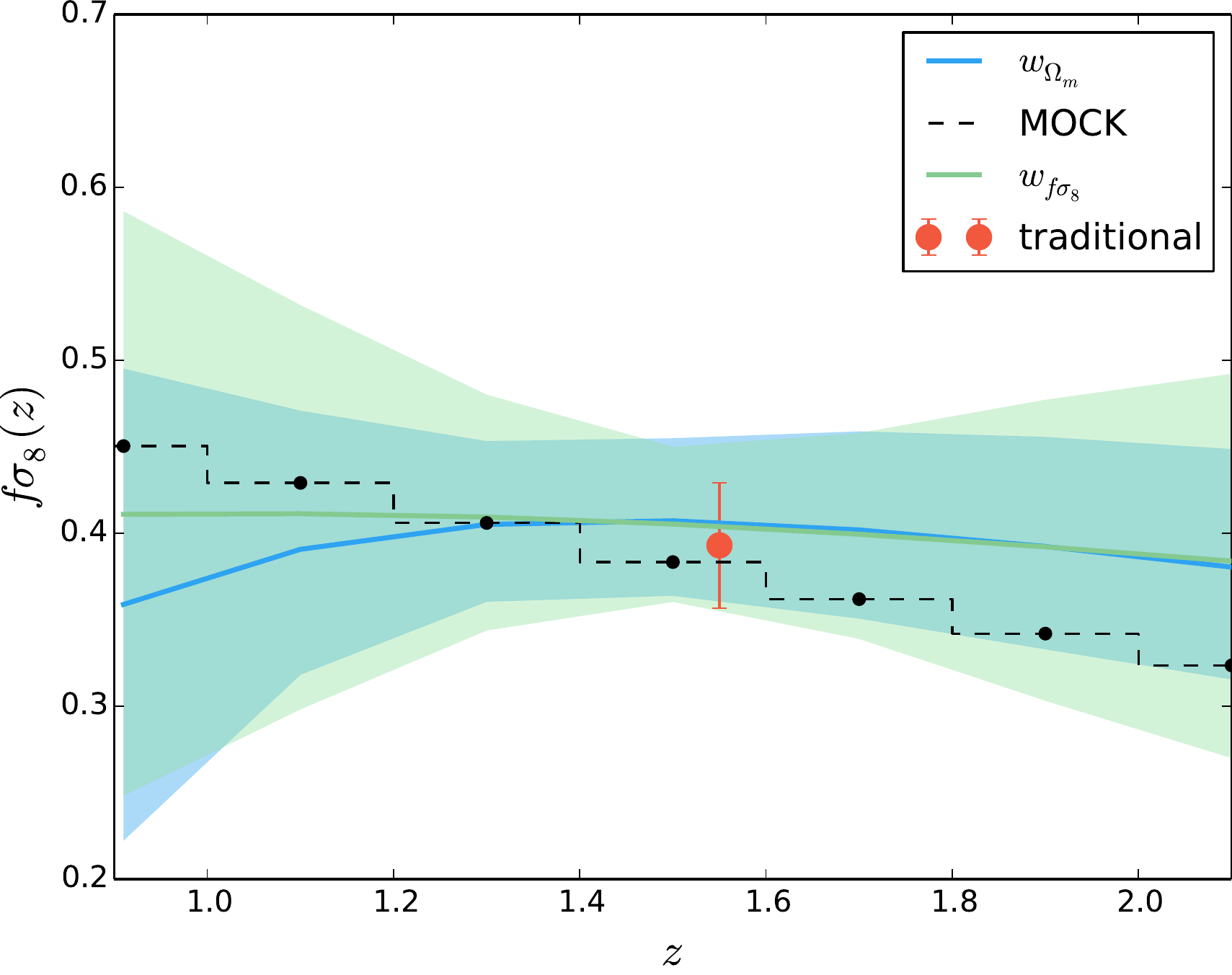}
\caption{The reconstructed evolution of $f\sigma_8$ and $68\% $ confidence level regions using the average of $20$ mocks;  blue shaded region shows the constraint on the evolution of $f \sigma_8$ obtained by the fit of $\Omega_m(z, q_{i}) $ using the $w_{\Omega_m}$ optimal weights and deriving at each redshift $f \big[\Omega_m(z, q_i) \big]$ times $\sigma_8 \big [\Omega_m (z, q_i) \big]$;  green shaded region shows the resulting evolution when fitting for $f\sigma_8(z, p_i)$ at each redshift. The red point indicates the results obtained when performing the traditional analysis, with $z_{\rm piv} = 1.55$. }
\label{sh_plotc}
\end{figure}

  Fig. \ref{sh_plotc} shows the redshift evolution reconstructed from  $p_0$, $p_1$, (green shaded regions), compared with the evolution reconstructed from the  $q_0$, $q_1$ (blue shaded regions). The red point indicates the constraints at one single redshift (traditional analysis, with $z$ = 1.55) for $f\sigma_8$. 
  We overplot the evolution of  $f\sigma_8(z)$ as accounted  in the mock lightcone (black dashed line).  
The plot shows that  the $f \sigma_8$ evolution obtained for both the $\Omega_m$ and $f\sigma_8$ weighting schemes  is fully consistent with the cosmology contained in the mock and in full agreement with the constraints coming from the traditional analyis.
  For both parametrizations the errors obtained at the pivot redshift is comparable with the error we get from  the traditional analysis. Note that the error from redshift weigthing analysis comes from the marginalization over a set of $7$ parameters in constrast to the traditional analyisis limited to only $5$ free parameters. 
 
  Away from from the pivot redshift, the errors becomes larger for both parametrizations. At these redshifts, the major contribution to the error comes from the slope constraints  ($q_1, p_1$) and the S/N is lowered due to the low number density $n(z)$, \citep{2017ata}.  For both parametrizations, the slope parameters are degenarate with the non linear bias parameters.

\begin{figure}
\includegraphics[scale=0.4]{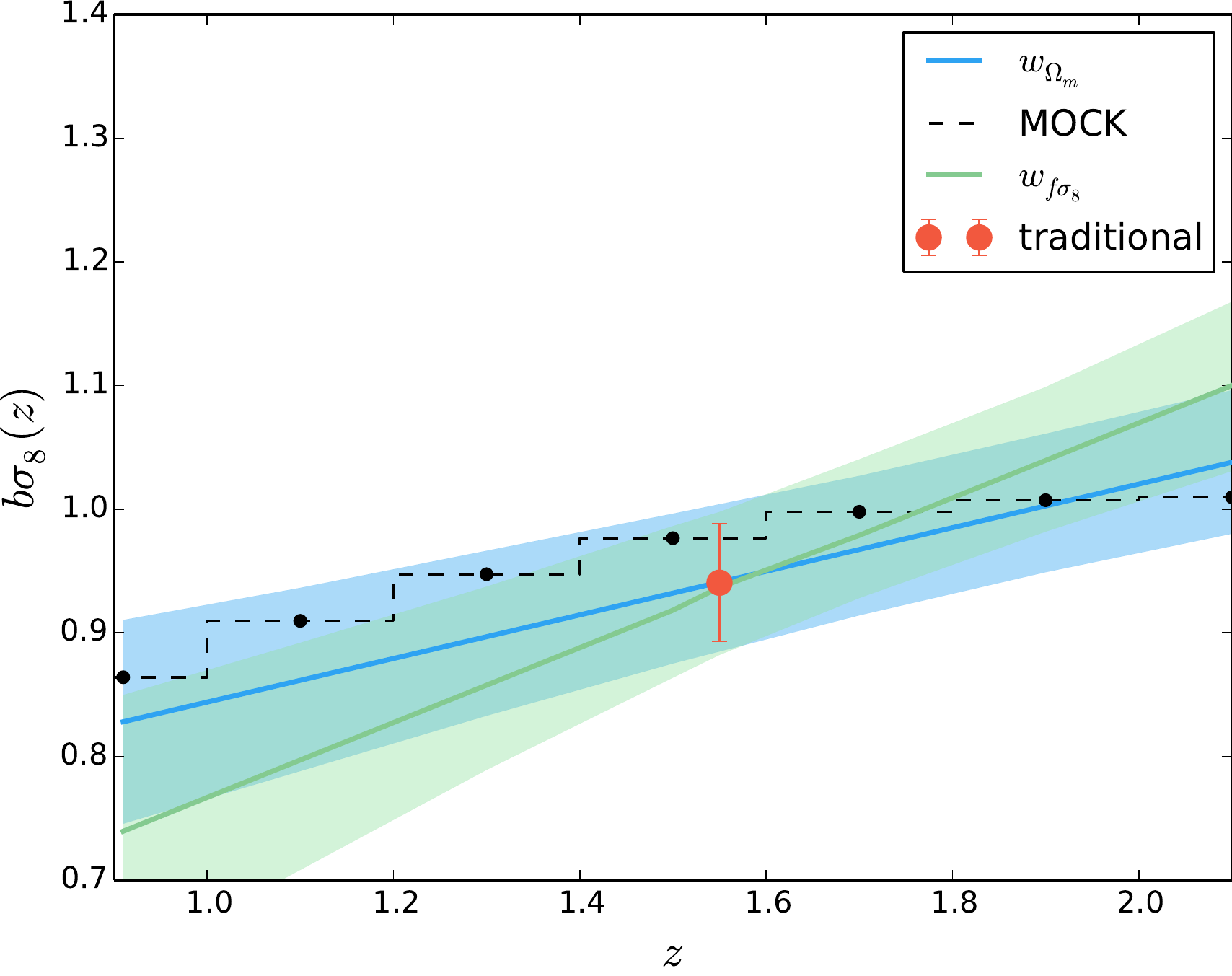}
\caption{The reconstructed evolution for $b\sigma_8 (z)$ and $68\% $ confidence level regions using the average of $20$ mocks;  we fit the evolution for $b\sigma_8$, modelled as a Taylor expansion about the pivot redshift, up to linear order. Blue shaded regions show the evolution of $b \sigma_8$ through the fit of $b\sigma_8(z_p)$, $\partial b \sigma_8(z) /\partial z $, obtained for the $\Omega_m(q_i)$ analysis;  green shaded regions show the analogous resulting $b\sigma_8 (z)$  when fitting for $f\sigma_8(z, p_i)$ at each redshift. The red point indicates the results obtained for $f\sigma_8(z_{piv})$ when performing the traditional analysis.
}
\label{sh_plotb}
\end{figure}

 In \ref{aa} we modelled the bias evolution with a Taylor expansion up to linear order about the pivot redshift (see Eq. \ref{biasevol}).   
  %
Fig. \ref{sh_plotb} shows the $b\sigma_8(z)$ evolution measured using  the $\Omega_m$ and $f\sigma_8$ weighting schemes (blue and green shaded regions). We reconstruct $b\sigma_8(z)$ at the different redshifts  from the fit of $b\sigma_8(z_p)$ and $\partial b \sigma_8(z) /\partial z $.  We overplot the evolution of  $b\sigma_8(z)$ as included  in the mocks (black dashed line). The red point indicates the constraints obtained by using the traditional analysis;
 we find full agreement at the pivot redshift between the three different analysis and within $1\sigma$ of the value included in the mocks. The bias 
 depends significantly on redshift and in the mocks is modelled as a step function, which leads to small discrepancies with respect to both  
 the constant and linear  evolution in $b\sigma_8$.
We redid the fit extending the analysis to second order in bias and found consistent results but with error too large to see any improvements (high degeneracy).
For the purpose of fitting  eBOSS quasar sample this is more than enough and  we leave for future work  a more careful study of the bias effects /evolution to be performed on more accurate nbody mocks. This is discussed further in section \ref{discuss} 
 %
 %
%

 \subsection{Constant bias vs evolving bias}\label{bb}
 We now  investigate how a particular choice for the bias evolution in redshift can affect and impact the constraints on $f\sigma_8(z)$. 
To do this, we repeat the   analysis as  presented in \ref{aa} using the $\Omega_m$ and $f\sigma_8$ weights, we model $\Omega_m(z)$ and $f\sigma_8(z)$ in the same way as in \ref{aa}, but now assuming that  the bias is constant with redshift i.e we set $\partial b \sigma_8(z) /\partial z  = 0 $. 
  
\begin{figure}
\includegraphics[scale=0.35]{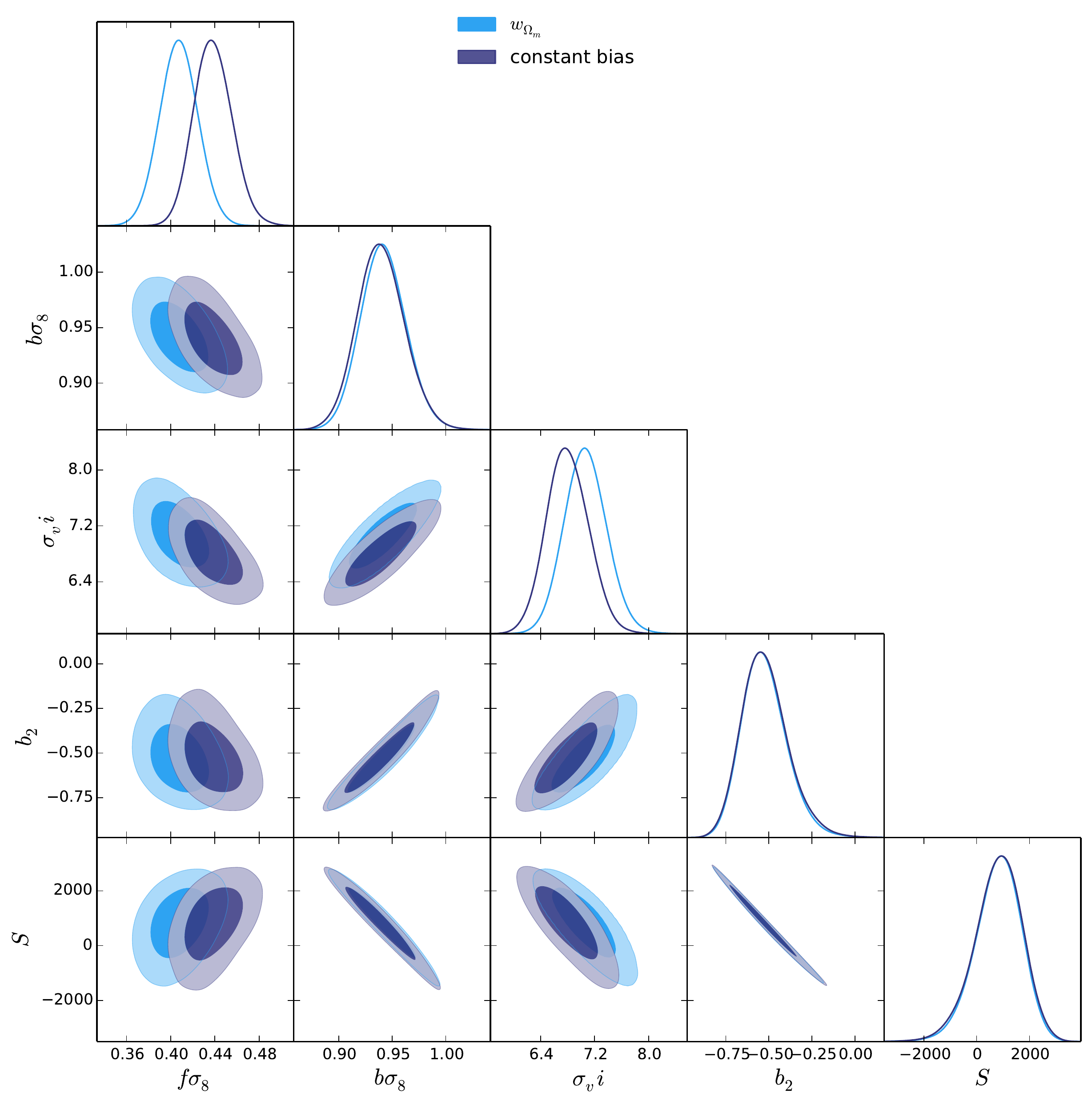}
\caption{Comparison between evolving and constant bias for the $\Omega_m$ - weights analysis.  
Blue likelihood contours indicate the constraints obtained when fitting for $b\sigma_8(z_p)$ and $\partial b \sigma_8(z) /\partial z $; 
dark blue contours indicate the constraints obtained when setting $\partial b \sigma_8(z) /\partial z  = 0 $ and fitting only for $b\sigma_8(z_p)$.}
\label{figura6}
\end{figure}

\begin{figure}
\includegraphics[scale=0.35]{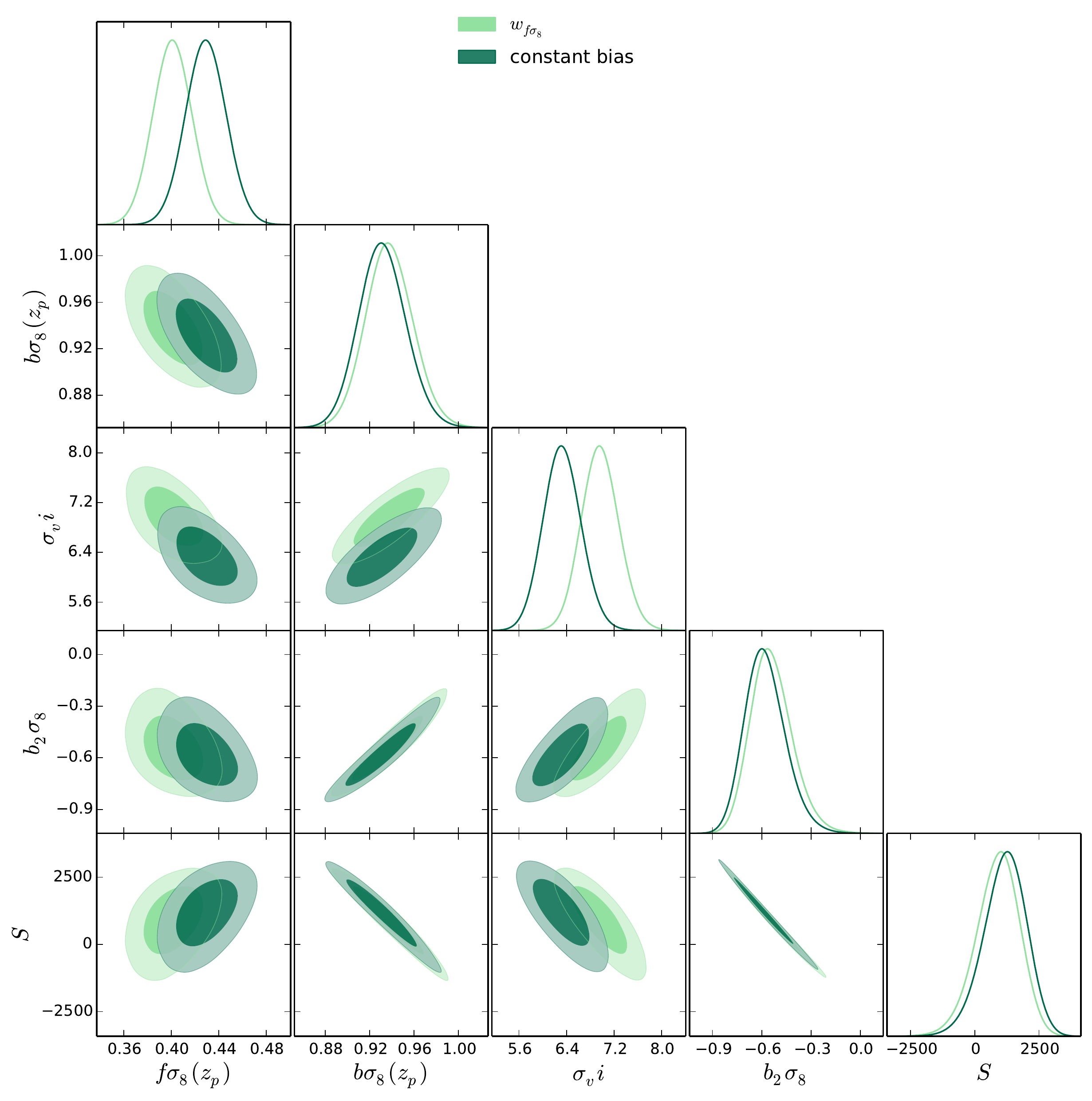}
\caption{Comparison between evolving and constant bias for the $f \sigma_8$-weights analysis.  
 Green likelihood contours indicate the constraints obtained when fitting for $b\sigma_8(z_p)$ and $\partial b \sigma_8(z) /\partial z $; 
dark green contours indicate the constraints obtained when setting $\partial b \sigma_8(z) /\partial z  = 0 $ and fitting only for $b\sigma_8(z_p)$.}
\label{figura6bis}
\end{figure}

 In Figures \ref{figura6}, \ref{figura6bis} we show the comparison between the results obtained with the constant bias.
 We display the posterior likelihood for all the quantities evaluated at the pivot redshift, $f\sigma_8(z_p)$ $b\sigma_8(z_p)$, $\sigma_v$, $b_2$, $S$. 
%
%
In Figure \ref{figura6},  blue contours show the likelihood distributions obtained when using the $\Omega_m$ weights and considering $b\sigma_8$ evolving as in Eq. \ref{biasevol}.  Dark blue contours indicate the constraints obtained when considering $\partial b \sigma_8(z) /\partial z  = 0 $.  In Figure  \ref{figura6bis} we present the analogous results when using the $f\sigma_8$ parametrization; green contours show the likelihood distributions obtained when using the $f\sigma_8$ weights considering the bias evolving as in Eq. \ref{biasevol}. Dark green contours correspond to the constraints obtained  when we set  $\partial b \sigma_8(z) /\partial z$ equal to zero. 
 The results obtained from the different models are consistent, but, whereas the constraints for $b\sigma_8(z_p)$ remain unchainged  there is an evident impact on the $f\sigma_8$ constraints at the pivot redshift. Forcing the bias to be constant with redshift lead to an higher value for $f\sigma_8$.
This should  be kept in mind for future work when higher precision is expected: a careful study/treatment of the bias is required in order to be sure not to bias the constraints on the growth.

 %
  %
 %

%
%
\subsection{Weights vs no Weights }\label{cc}
We compare the analysis performed using the redshift weights approach, as presented in \ref{aa}  with the \textit{traditional} analysis at one constant redshift. 

  %
The traditional analysis makes use of the power spectrum moments, modelled as in Sec. \ref{onez}, to constrain $f\sigma_8$ and $b\sigma_8$ at one single epoch which corresponds to the effective redshift of the survey ($z=1.55$). 
We do the comparison for both the $\Omega_m$  $f\sigma_8$ weigthing schemes; 

\begin{figure}
\includegraphics[scale=0.35]{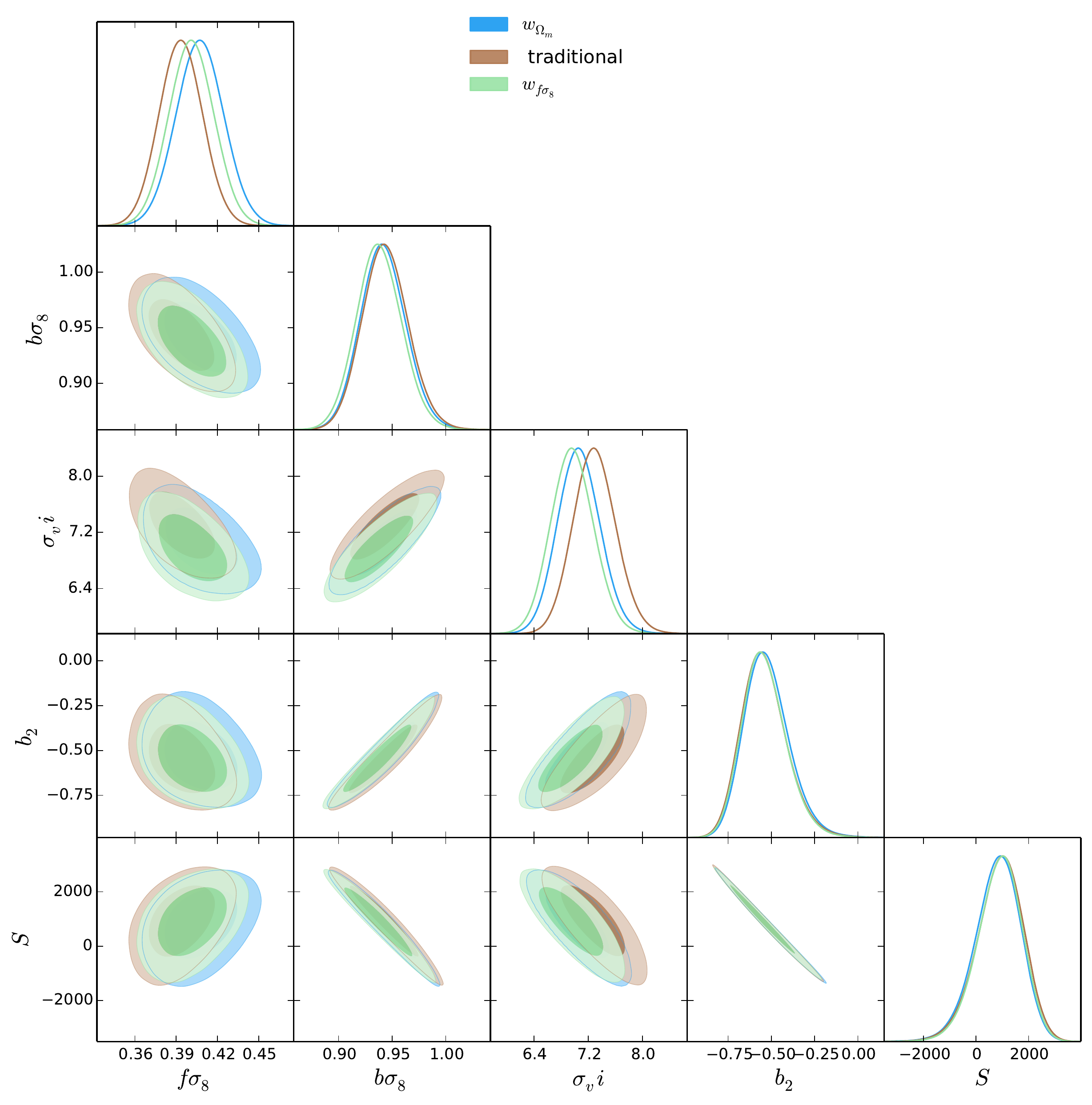}
\caption{The comparison between the redshift weights analysis and the traditional analysis. Likelihood contours for $f\sigma_8$, $b\sigma_8$, $\sigma_v$ $b_2$, $S$ quantities, at their pivot redshift values. 
Blue likelihood contours show the results obtained with the $\Omega_m(q_i)$ analysis; green contours show the results from the $f\sigma_8(p_i)$ analysis. Brown contours indicate the results obtained with the traditional analyisis.  }
\label{figura5}
\end{figure}

Figure \ref{figura5} shows the comparison between the redshift weights analysis for $\Omega_m$ (blue contours), $f\sigma_8$ (green contours) and the constant redshift analysis (brown contours). 
 In order to make the comparison between the three different analysis we infer from the MCMC chains of $q_i$ and $p_i$,    the $f\sigma_8[z, \Omega_m(q_i)]$ and  $f\sigma_8(z, p_i)$ valued at the pivot redshifts. We then compare those values with the $f\sigma_8(z_p)$, $b\sigma_8(z_p)$ as obtained from the traditional analysis.
 The last two panels in Fig. \ref{figura5} show that we recover the same value for $b_2$, and $S$ where the evolution in redshift is not considered in all the three different analysis;  the other constraints on $f\sigma_8$, $b\sigma_8$ and $\sigma_v$ are fully consistent within $\sim 1 \sigma$. 
   %
  %

%
%
%
%

\section{Conclusion}\label{discuss}
In the current work we present a new approach to measure redshift space distortions when dealing with surveys covering a wide redshift range; 
the redshift weights, applied to each galaxy within the sample, act as a smooth window on the data, allowing us to compress the information in the redshift direction without  loss of information. In this analysis we applied the redshift weighting technique to investigate small deviation from the $\Lambda$CDM framework; we selected two different parametrization, allowing for deviation in the matter energy density and the growth rate evolution. We derived multiple  sets of weights to optimize each order of those deviations. 
We extended the window function derivation in order to account for the redshift evolution of the power spectrum. 

We compared the results obtained for the different parameters  with the traditional analysis, i.e. the analysis perfomed considering the clustering as constant in the whole volume. We found that the redshift weights technique gives unbiased constraints for the whole redshift range, in full agreement with the traditional analysis performed at the effective redshift.

The constraints obtained fully validate the analysis (Ruggeri et al. 2017 in prep) to measure RSD on the eBOSS quasars sample where the error expected on $f\sigma_8$ is about $5\%$.  To apply the same pipeline to future surveys aiming at $\%$ level accuracy further work will be required; firstly we will need to consider quadratic deviations in the evolution for both the $q_i$ and the galaxy bias parameters. In this work we only accounted for those deviations to test the robustness of the fits
whereas the signal expected from the quasars sample will not be able to constrain  the quadratic evolution. 

Another important and interesting aspect would be to account for the AP parameters and their evolution in redshift. 
To perform such analysis, a  set of $N-\rm body$ simulations that accurately describe non linearities/light-cone evolution is also required,  to reduce the degeneracies and provide a lower statistical error. 
   We here only considered   the growth alone, with better data we would be able to include both AP and growth. For the eBOSS sample, the constraints are too weak to consider this. 

\section*{Acknowledgements}
RR thanks Dr. VP, Dr. RR and Dr. GP for all the support provided.
RR and WJP acknowledges support from the European Research Council through
the Darksurvey grant 614030, from the UK Science and Technology
Facilities Council grant ST/N000668/1,   
 WJP acknowledges  the UK Space Agency grant
ST/N00180X/1.

Funding for SDSS-III and SDSS-IV has been provided by the Alfred
P. Sloan Foundation and Participating Institutions. Additional funding
for SDSS-III comes from the National Science Foundation and the
U.S. Department of Energy Office of Science. Further information about
both projects is available at www.sdss. org. SDSS is managed by the
Astrophysical Research Consortium for the Participating Institutions
in both collaborations. In SDSS- III these include the University of
Arizona, the Brazilian Participation Group, Brookhaven National
Laboratory, Carnegie Mellon University, University of Florida, the
French Participation Group, the German Participation Group, Harvard
University, the Instituto de Astrofisica de Canarias, the Michigan
State / Notre Dame / JINA Participation Group, Johns Hopkins
University, Lawrence Berkeley National Laboratory, Max Planck
Institute for Astrophysics, Max Planck Institute for Extraterrestrial
Physics, New Mexico State University, New York University, Ohio State
University, Pennsylvania State University, University of Portsmouth,
Princeton University, the Spanish Participation Group, University of
Tokyo, University of Utah, Vanderbilt University, University of
Virginia, University of Washington, and Yale University.  The
Participating Institutions in SDSS-IV are Carnegie Mellon University,
Colorado University, Boulder, Harvard- Smithsonian Center for
Astrophysics Participation Group, Johns Hopkins University, Kavli
Institute for the Physics and Mathematics of the Universe
Max-Planck-Institut fuer Astrophysik (MPA Garching),
Max-Planck-Institut fuer Extraterrestrische Physik (MPE),
Max-Planck-Institut fuer Astronomie (MPIA Heidelberg), National
Astronomical Observatories of China, New Mexico State University, New
York University, The Ohio State University, Penn State University,
Shanghai Astronomical Observatory, United Kingdom Participation Group,
University of Portsmouth, Univer- sity of Utah, University of
Wisconsin, and Yale University.  This work made use of the facilities
and staff of the UK Sciama High Performance Computing cluster
supported by the ICG, SEPNet and the University of Portsmouth. This
research used resources of the National Energy Research Scientific
Computing Center, a DOE Office of Science User Facility supported by
the Office of Science of the U.S. Department of Energy under Contract
No. DE-AC02-05CH11231.

%
%
%
%


\def\jnl@style{\it}
\def\aaref@jnl#1{{\jnl@style#1}}

\def\aaref@jnl#1{{\jnl@style#1}}

\def\aj{\aaref@jnl{AJ}}                   
\def\araa{\aaref@jnl{ARA\&A}}             
\def\apj{\aaref@jnl{ApJ}}                 
\def\apjl{\aaref@jnl{ApJ}}                
\def\apjs{\aaref@jnl{ApJS}}               
\def\ao{\aaref@jnl{Appl.~Opt.}}           
\def\apss{\aaref@jnl{Ap\&SS}}             
\def\aap{\aaref@jnl{A\&A}}                
\def\aapr{\aaref@jnl{A\&A~Rev.}}          
\def\aaps{\aaref@jnl{A\&AS}}              
\def\azh{\aaref@jnl{AZh}}                 
\def\baas{\aaref@jnl{BAAS}}               
\def\jrasc{\aaref@jnl{JRASC}}             
\def\memras{\aaref@jnl{MmRAS}}            
\def\mnras{\aaref@jnl{MNRAS}}             
\def\pra{\aaref@jnl{Phys.~Rev.~A}}        
\def\prb{\aaref@jnl{Phys.~Rev.~B}}        
\def\prc{\aaref@jnl{Phys.~Rev.~C}}        
\def\prd{\aaref@jnl{Phys.~Rev.~D}}        
\def\pre{\aaref@jnl{Phys.~Rev.~E}}        
\def\prl{\aaref@jnl{Phys.~Rev.~Lett.}}    
\def\pasp{\aaref@jnl{PASP}}               
\def\pasj{\aaref@jnl{PASJ}}               
\def\qjras{\aaref@jnl{QJRAS}}             
\def\skytel{\aaref@jnl{S\&T}}             
\def\solphys{\aaref@jnl{Sol.~Phys.}}      
\def\sovast{\aaref@jnl{Soviet~Ast.}}      
\def\ssr{\aaref@jnl{Space~Sci.~Rev.}}     
\def\zap{\aaref@jnl{ZAp}}                 
\def\nat{\aaref@jnl{Nature}}              
\def\iaucirc{\aaref@jnl{IAU~Circ.}}       
\def\aplett{\aaref@jnl{Astrophys.~Lett.}} 
\def\apspr{\aaref@jnl{Astrophys.~Space~Phys.~Res.}}
\def\bain{\aaref@jnl{Bull.~Astron.~Inst.~Netherlands}} 
\def\fcp{\aaref@jnl{Fund.~Cosmic~Phys.}}  
\def\gca{\aaref@jnl{Geochim.~Cosmochim.~Acta}}   
\def\grl{\aaref@jnl{Geophys.~Res.~Lett.}} 
\def\jcp{\aaref@jnl{J.~Chem.~Phys.}}      
\def\jgr{\aaref@jnl{J.~Geophys.~Res.}}    
\def\jqsrt{\aaref@jnl{J.~Quant.~Spec.~Radiat.~Transf.}}
\def\memsai{\aaref@jnl{Mem.~Soc.~Astron.~Italiana}}
\def\nphysa{\aaref@jnl{Nucl.~Phys.~A}}   
\def\physrep{\aaref@jnl{Phys.~Rep.}}   
\def\physscr{\aaref@jnl{Phys.~Scr}}   
\def\planss{\aaref@jnl{Planet.~Space~Sci.}}   
\def\procspie{\aaref@jnl{Proc.~SPIE}}   
\def\jcap{\aaref@jnl{J. Cosmology Astropart. Phys.}}

\let\astap=\aap
\let\apjlett=\apjl
\let\apjsupp=\apjs
\let\applopt=\ao

\newcommand{\mpc}{\, {\rm Mpc}}
\newcommand{\kpc}{\, {\rm kpc}}
\newcommand{\hmpc}{\, h^{-1} \mpc}
\newcommand{\ihmpc}{\, h\, {\rm Mpc}^{-1}}
\newcommand{\ikms}{\, {\rm s\, km}^{-1}}
\newcommand{\kms}{\, {\rm km\, s}^{-1}}
\newcommand{\hkpc}{\, h^{-1} \kpc}
\newcommand{\lya}{Ly$\alpha$\ }
\newcommand{\lyb}{Lyman-$\beta$\ }
\newcommand{\lyaf}{Ly$\alpha$ forest}
\newcommand{\lr}{\lambda_{{\rm rest}}}
\newcommand{\bF}{\bar{F}}
\newcommand{\bS}{\bar{S}}
\newcommand{\bC}{\bar{C}}
\newcommand{\bB}{\bar{B}}
\newcommand{\vdF}{{\mathbf \delta_F}}
\newcommand{\vdS}{{\mathbf \delta_S}}
\newcommand{\vdf}{{\mathbf \delta_f}}
\newcommand{\vdn}{{\mathbf \delta_n}}
\newcommand{\vdC}{{\mathbf \delta_C}}
\newcommand{\vdX}{{\mathbf \delta_X}}
\newcommand{\xrei}{x_{rei}}
\newcommand{\lrmin}{\lambda_{{\rm rest, min}}}
\newcommand{\lrmax}{\lambda_{{\rm rest, max}}}
\newcommand{\lmin}{\lambda_{{\rm min}}}
\newcommand{\lmax}{\lambda_{{\rm max}}}
\newcommand{\hi}{\mbox{H\,{\scriptsize I}\ }}
\newcommand{\heii}{\mbox{He\,{\scriptsize II}\ }}
\newcommand{\vp}{\mathbf{p}}
\newcommand{\vq}{\mathbf{q}}
\newcommand{\vxperp}{\mathbf{x_\perp}}
\newcommand{\vkperp}{\mathbf{k_\perp}}
\newcommand{\vrperp}{\mathbf{r_\perp}}
\newcommand{\vx}{\mathbf{x}}
\newcommand{\vy}{\mathbf{y}}
\newcommand{\vk}{\mathbf{k}}
\newcommand{\vR}{\mathbf{r}}
\newcommand{\tdtwo}{\tilde{b}_{\delta^2}}
\newcommand{\tstwo}{\tilde{b}_{s^2}}
\newcommand{\tbthree}{\tilde{b}_3}
\newcommand{\tadtwo}{\tilde{a}_{\delta^2}}
\newcommand{\tastwo}{\tilde{a}_{s^2}}
\newcommand{\tabthree}{\tilde{a}_3}
\newcommand{\vnabla}{\mathbf{\nabla}}
\newcommand{\tpsi}{\tilde{\psi}}
\newcommand{\vv}{\mathbf{v}}
\newcommand{\fnl}{{f_{\rm NL}}}
\newcommand{\tfnl}{{\tilde{f}_{\rm NL}}}
\newcommand{\gnl}{g_{\rm NL}}
\newcommand{\orderfour}{\mathcal{O}\left(\delta_1^4\right)}
\newcommand{\SDSSPF}{\cite{2006ApJS..163...80M}}
\newcommand{\PF}{$P_F^{\rm 1D}(k_\parallel,z)$}
\newcommand\ionalt[2]{#1$\;${\scriptsize \uppercase\expandafter{\romannumeral #2}}}%
\newcommand{\vxone}{\mathbf{x_1}}
\newcommand{\vxtwo}{\mathbf{x_2}}
\newcommand{\vRot}{\mathbf{r_{12}}}
\newcommand{\cm}{\, {\rm cm}}

\bibliography{draft}

\label{lastpage}

\newpage

\appendix

\end{document}